## Python GUI Scripting Interface for Running Atomic Physics Applications

Amani Tahat[1, 2, 6], Mofleh Tahat[3], Wa'el Salah[4, 5]

[1]Department of Basic Science, Philadelphia University, Amman, Jordan

[3]Jordan University of Science and Technology, Irbid, Jordan

[2,4]Physics Department, The Hashemite University, Zarqa, 13133, Jordan

[5]Synchrotron-light for Experimental Science and Applications in the Middle East (SESAME), Jordan

[6]Email: amanitahat@yahoo.com,

**Keywords:** Graphical user interface, Atomic Calculations

**Abstract**

We create a Python GUI scripting interface working under Windows in addition to (UNIX/ Linux). The GUI has been built around the Python open-source programming language. We use the Python's GUI library that so called Python Mega Widgets (PMW) and based on Tkinter Python module (http://www.freenetpages.co.uk/hp/alan.gauld/tutgui.htm). The new GUI was motivated primarily by the desire of more updated operations, more flexibility incorporating future and current improvements in producing atomic data. Furthermore it will be useful for a variety of applications of atomic physics, plasma physics and astrophysics and will help in calculating various atomic properties.

## 1.    Introduction

This GUI development has been particularly established for teaching purposes and has arisen as a result of demand within the Jordanian academic community. We plan to satisfy using an extensible Graphical User Interface (GUI) for atomic codes.

The goal of this work focuses on creating a simple and highly flexible GUI system that allows easy creation of dialogs for performing the calculations of large range of atomic data; based on some important production software that mainly written in Python, such as the open source atomic applications FAC (Flexible Atomic Code) (Gu, 2008). Currently this GUI is created for running the FAC functions exclusively, but it could be suitable to be wired into another application by changing its form. More descriptions about building this GUI are available in (Tahat et al., 2010).

This manuscript presents a Python source code of a flexible scripting user interface to work as a utility program for the atomic package FAC. Using this GUI will make it easy to determine the input and schedule the computations, in addition to allow typical calculations to be carried out with minimal understanding of theoretical atomic physics. Furthermore, it allows saving the output at any specified directory.

The layout of the paper is as follows: In the next section we summarize the main ideas of our approach along with a short demonstration of using the Flexible Atomic Code (FAC), followed by the way of running the scripting GUI. The paper ends with the source code file.

The FAC code is a powerful atomic code; it handles atomic calculations, collisional radiative spectral models, and is used to calculate line polarizations due to collisional excitation. It is a flexible atomic code and can work as an integrated software package for the calculation of a variety of atomic properties.

FAC provides two interfaces to run the code: SFAC and PFAC. Furthermore, when using the SFAC interface, users are typically unable to customize or modify features further than cosmetic aspects. Although in FAC calculations there are many important input parameters that a user should be aware of, and a mixture of computing models that are based on different physical assumptions that can be activated by using the PFAC interface. The current GUI has been designed to work as new FAC interface to be easy to run and to produce data files that can interface with other programs easily. Moreover, GUI scripting helps users to specify the problem, interact with the program dynamically and also help users analyze the output. For more information the reader should refer to the FAC manual (Gu, 2003) to understand FAC functions and input parameters.

Step-by-step instructions for setting up SFAC and PFAC interfaces can be found in the README file of the FAC package and can be used for comparing the installation process of the current GUI with those two interfaces of FAC. Moreover, much of the FAC package is written in ANSI C and Fortran 77. It should therefore work on any platform with a C and Fortran 77 compilers. On the other hand, this is only true to the rather simple command parser that comes with FAC, referred to as SFAC .Additionally, for using PFAC it is strongly recommended that Python be installed. In the case of Windows, the Unix API emulation by Cygwin is required, which is available at www.cygwin.com.

In particular to fully utilize the strength of the FAC code, this GUI allows running all parts of PFAC in a very simple way without installing any application or programming language, regardless of the operating system, whether it is being Windows or UNIX/Linux.

For using this GUI user has to follow the instructions of using this new interface, starting from calling the input file of the FAC code based on the part that he wishes. The last step must be followed by pressing the calculate button; in order to produce the output file. In the meantime user can save the output file after viewing it in any directory of his computer.

This scripting GUI will be the first step of building a full option GUI. The GUI scripting will help a programmer when consulting physicists during designing the GUI interface (e.g. what icons, menu, box, counter, tabs, etc.) does a user need to compose the input window in order to facilitate creating the input file for each model in FAC code. In addition, it is a simple GUI that enables non-IT people to conduct the computations easily. This plays a good role in developing application in terms of Interfaces adaptability. A manuscript published in The Python Papers (TPP) (Tahat, 2010) presents a good example of creating such features based on the current GUI as a first step for design the full option GUI.

Therefore, this work can be considered as a successful methodology for the creation of adaptable user interfaces by using the Python's GUI library that specifically called Python Mega Widgets (PMW) (http://pmw.sourceforge.net) based on the Tkinter Python module (Grayson, 2000). Figure 1 presents the structure of building a powerful system for running the FAC code by using the current GUI as well as summarizing the way of building this GUI.

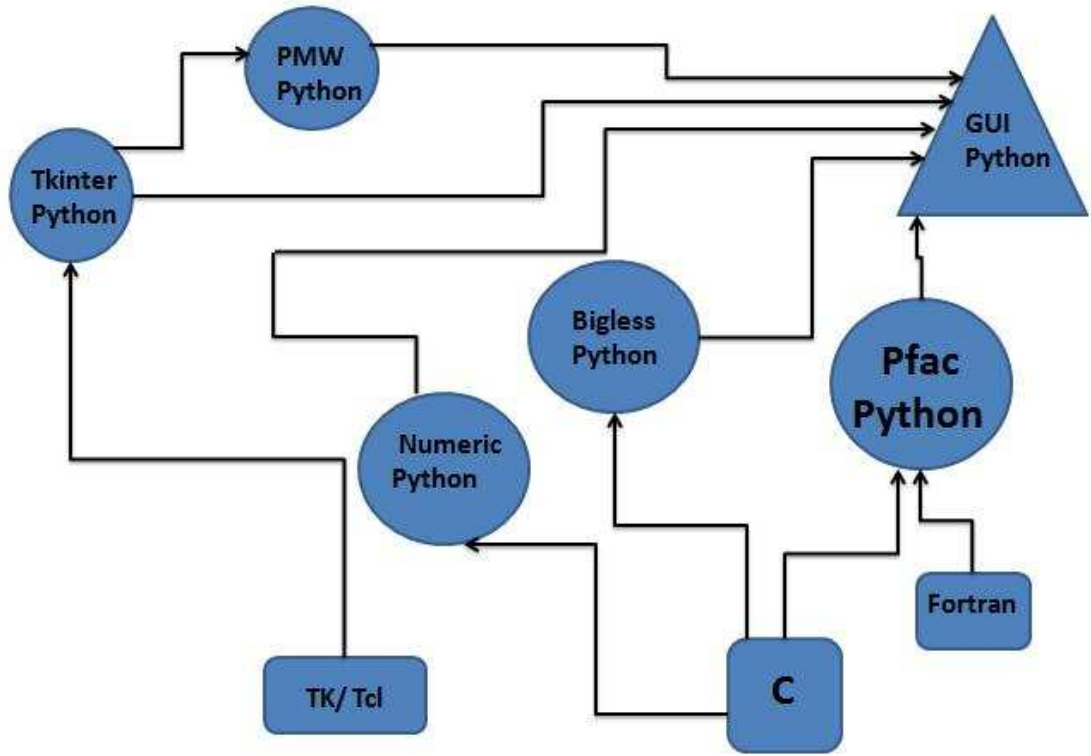

Figure1: The interconnection of the current GUI and PFAC along with their programming languages. Much of the PFAC functions is written in C and Fortran. The Biggles Python model (http://biggles.sourceforge.net/.) can be added to allow plotting capabilities as an advanced feature for the current GUI in order to illustrate the atomic data that tabulated in the output files. The numeric Python module (http://people.csail.mit.edu/jrennie/python/numeric/) will allow defining the multidimensional array and useful procedures for numerical computation of PFAC functions.

## 2.      Running the Scripting GUI

The scripting window consists of two parts: The input panel and output panel. The user enters scripts in the input part and clicks the "calculate" button. The Python script engine processes the script and the result is written to the output part. The output/input can be saved by choosing "file menu -> save". Moreover, to make working with scripts more convenient, it is possible to set up directories where scripts maybe stored. The feature is available from the "file menu -> open". This file can be edited and saved, in the meantime, and all files will be stored in the output directory which can be set by the user.

The following Python scripts refer to particular piece of the source code that enable saving the output results as a text file.

```
def save():
    """Save result."""
    content = text.get('1.0', END)
    try:
        fd = open(fname.text, 'w')
        fd.write(content)
        fd.close()
    except IOError, m:
        showerror("Write File...", m)
```

In the meantime, we are trying to present the way of adding some utility features to any Python GUI like save, view, delete, along with print, etc.

This manuscript will work as a helpful beginner's user guide to Python GUI scripting interface. The following example presents the way of writing the Python scripts for viewing the output results:

```
def view():
    """View results."""
    try:
        the_file = askopenfilename(filetypes=[("all files", "*")])
        if the_file == "":
            return
        fd = open(the_file, 'r')
        lines = fd.read()
        fd.close()
        view.delete('1.0', END)
        view.insert(END, lines)
    except IOError, m:
        showerror("Open File...", m)
```

Furthermore, some attention massages will be appeared while running this GUI. For example, a waiting massage will be appeared during the calculations. A new massage will pop up for announcing the end of calculation process as well. The following Python scripts show the easy creation of such massages:

```
def run():
    """Run FAC script."""
    info.configure(text="Calculation... wait...")
    info.update()
    exec text.get('1.0', END)
    view.delete('1.0', END)
    info.configure(text="")
    showinfo("Work done", "Please see results in folder files")
    fac.Reinit(0)
```

Figure 2 presents a screenshot of the main panel of the scripting window including an attention massage that announcing the end of calculation process.

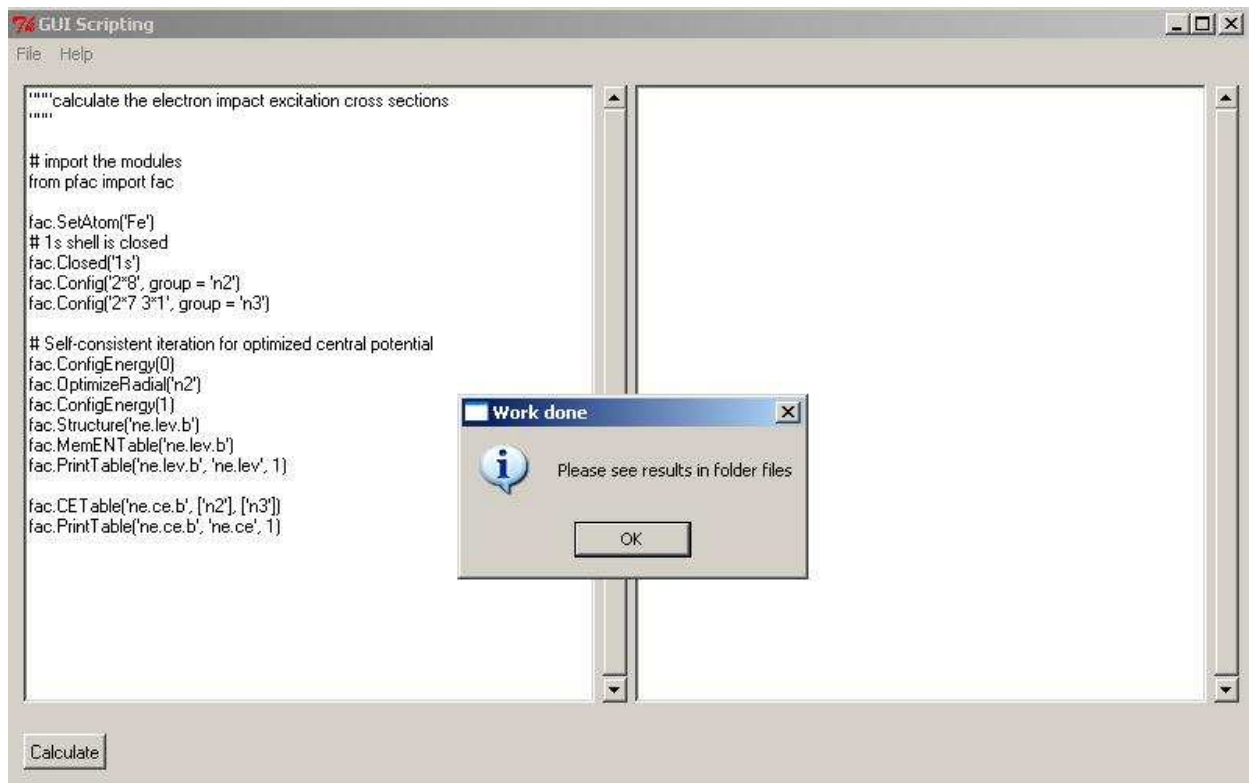

Figure 2: The GUI scripting interface with its two panels (input/output). The input panel is filled with FAC scripts for calculating the electron impact excitation cross section.

Finally this work is an attempt to provide a guide in using Tkinter programming aspects that can help easily in putting a simple GUI front-end on Python applications. Moreover the produced GUI will be useful in generating the following atomic data: radiative transition rates, cross sections of electron impact excitation along with ionization, photoionization, autoionization, as well as their inverse processes, radiative and dielectronic recombination.

## 3.     Code file

This GUI has been successfully used in producing atomic data for some master thesis in several Jordanian universities besides using it in a number of important atomic researches. This GUI has been released under the Lesser General Public License version 3 (LGPL 3) and is freely distributed.

```
import os
from tkMessageBox import *
from tkFileDialog import *
from Tkinter import *
from pfac import fac

about = 'Produced by: AMANI NASER TAHAT '
```

```python
def load():
    """Load FAC script."""
    try:
        the_file = askopenfilename(filetypes=[("python files", "*.py"),
                                              ("all files", "*")],
                                   initialdir="./files")
        if the_file == "":
            return
        fname.text = the_file
        fd = open(the_file, 'r')
        lines = fd.read()
        fd.close()
        text.delete('1.0', END)
        text.insert(END, lines)
        calc.configure(state=NORMAL)
        calc.flash()
    except IOError, m:
        showerror("Open File...", m)

def view():
    """View results."""
    try:
        the_file = askopenfilename(filetypes=[("all files", "*")])
        if the_file == "":
            return
        fd = open(the_file, 'r')
        lines = fd.read()
        fd.close()
        view.delete('1.0', END)
        view.insert(END, lines)
    except IOError, m:
        showerror("Open File...", m)

def save():
    """Save result."""
    content = text.get('1.0', END)
    try:
        fd = open(fname.text, 'w')
        fd.write(content)
        fd.close()
    except IOError, m:
        showerror("Write File...", m)

def run():
    """Run FAC script."""
    info.configure(text="Calculation... wait...")
    info.update()
    exec text.get('1.0', END)
    view.delete('1.0', END)
    info.configure(text="")
    showinfo("Work done", "Please see results in folder files")
    fac.Reinit(0)

# Main programme with GUI loop.
```

```python
if __name__ == "__main__":
    os.chdir("./files")
    root = Tk()
    root.title("GUI Scripting")
    fname = Label()
    menu = Menu(root)
    root.config(menu=menu)
    file_menu = Menu(menu)
    menu.add_cascade(label="File", menu=file_menu)
    file_menu.add_command(label="Open", command=load)
    file_menu.add_command(label="Save", command=save)
    file_menu.add_command(label="View", command=view)
    file_menu.add_command(label="Exit", command=root.destroy)

    help_menu = Menu(menu)
    menu.add_cascade(label="Help", menu=help_menu)
    help_menu.add_command(label="About",
                          command=(lambda: winfo('About', about)))

    frame = Frame(root)
    text=Text(frame, height=30, width=60)
    text.pack(side=LEFT, fill=X, padx=5)

    sb = Scrollbar(frame, orient=VERTICAL, command=text.yview)
    sb.pack(side=LEFT, fill=Y)

    text.configure(yscrollcommand=sb.set)
    view=Text(frame, height=30, width=60)
    view.pack(side=LEFT, fill=X, padx=5)
    sb2 = Scrollbar(frame, orient=VERTICAL, command=view.yview)
    sb2.pack(side=RIGHT, fill=Y)

    view.configure(yscrollcommand=sb2.set)
    frame.pack(expand=1, fill=X, pady=10, padx=5)

    calc = Button(text='Calculate', command=run,state=DISABLED)
    calc.pack(side=LEFT, pady=10, padx=10)

    info = Label(text='')
    info.pack(side=LEFT, pady=10, padx=10)

    root.mainloop()
```

## References

Gu, M. F. 2008, Review /syntheses. The flexible atomic code, Canadian Journal of Physics. (86): 675-689.

Gu, M. F. 2003. FAC 1.1.0 Manual.

Grayson, J. 2000. Python and Tkinter Programming. Manning, ISBN 1-884777-81-3.

Tahat, A., Salah, W., Tahat, M. 2010. National library of Jordan. GUI scripting interface. ISBN (12/08/2010).

Tahat, A. 2010, An atomic multiplet code for the calculation of various atomic properties. The Python Papers, 5(1).